\documentclass[aps,reprint,showpacs,superscriptaddress]{revtex4-1}  
\usepackage{graphicx}  
\usepackage{dcolumn}   
\usepackage{bm}        
\usepackage{amssymb}   
\usepackage{amsmath}
\usepackage{slashed}
\usepackage{mathtools}
\usepackage{braket}
\usepackage{hyperref}
\usepackage{cleveref}
\usepackage{tikz-feynman}
\usepackage{dsfont}

\DeclarePairedDelimiter{\floor}{\lfloor}{\rfloor}

\allowdisplaybreaks

\newmuskip\pFqmuskip
\newcommand*\pFq[6][8]{%
  \begingroup 
  \pFqmuskip=#1mu\relax
  \mathchardef\normalcomma=\mathcode`,
  \mathcode`\,=\string"8000
  \begingroup\lccode`\~=`\,
  \lowercase{\endgroup\let~}\pFqcomma
  {}_{#2}{F}_{#3}{\bigg[\genfrac..{0pt}{}{#4}{#5};#6\bigg]}%
  \endgroup
}
\newcommand{\pFqcomma}{{\normalcomma}\mskip\pFqmuskip}

\newcommand{\cE}{\mathcal{E}}

\newcommand{\cM}{\mathcal{M}}
\newcommand{\dd}{\mathrm{d}}
\newcommand{\fa}{\mathfrak{a}}

\begin{document}

\hspace{5.2in} \mbox{CERN-TH-2025-183}
\preprint{CERN-TH-2025-183}

\title{Hidden simplicity in the scattering for neutron stars and black holes}
			     
\author{Rafael Aoude}                             
	
\email[]{rafael.aoude@ed.ac.uk}
	
\affiliation{Higgs Centre for Theoretical Physics,
	School of Physics and Astronomy,\\
	The University of Edinburgh, Edinburgh EH9 3JZ, Scotland, UK}%
			     
\author{Andreas Helset}                             

\email[]{andreas.helset@cern.ch}

\affiliation{Theoretical Physics Department, CERN, 1211 Geneva 23, Switzerland}

\date{\today}

\begin{abstract}
Heavy particle effective theory applied to spinning black holes provides a natural framework in which propagators linearize and numerators exponentiate. In this work, we exploit these two features to introduce Kerr generating functions, which describe the scattering of any probe on a Kerr black hole background to all loop orders. These generating functions can be used to perform the tensor reduction of multi-loop integrands simply by differentiation with respect to the spin. As a first application of the Kerr generating functions, we study the leading non-linear tidal effects of a neutron star in a Kerr black hole background. We organize the integrand by the helicity configuration of the exchanged gravitons and provide compact all-loop-order results for several helicity sectors and a full four-loop $\mathcal{O}(G^5)$ result for the leading non-linear tidal operators.
\end{abstract}

\maketitle


\section{Introduction}

Since the first observation of gravitational waves from binary coalescence, many new analytical methods have been developed to describe the two-body problem. Some are based on perturbative quantum field theory (QFT) techniques, where scattering amplitudes provide on-shell, gauge invariant, relativistic, and compact formulae encoding the dynamics of the two bodies. As is customary in perturbation theory, gravitational amplitudes are expanded in the coupling---Newton's constant $G$---while keeping exact dependence on the velocity. This perturbative series is called the Post-Minkowskian (PM) expansion. Progress in this direction leverages decades of knowledge from the multi-loop QCD precision frontier and has led to striking predictions in spinless scattering~\cite{Cheung:2018wkq,Bern:2019nnu,Bern:2021dqo,Bern:2021yeh,Bern:2024adl,Brammer:2025rqo}, radiation~\cite{Kosower:2018adc,Cristofoli:2021vyo,Herrmann:2021lqe,Herrmann:2021tct,Elkhidir:2023dco,Herderschee:2023fxh,Brandhuber:2023hhy,Georgoudis:2023lgf,Georgoudis:2023eke,Heissenberg:2025ocy,Georgoudis:2025vkk}, and tidal effects~\cite{Bern:2020uwk,Haddad:2020que,Aoude:2020ygw,Heissenberg:2022tsn,Mougiakakos:2022sic}. 
This development comes together with unprecedented progress in the wordline-based formalism~\cite{Kalin:2020mvi,Kalin:2020fhe,Liu:2021zxr,Dlapa:2021npj,Dlapa:2021vgp,Mougiakakos:2021ckm,Mougiakakos:2022sic,Riva:2022fru,Kalin:2022hph,Dlapa:2024cje,Dlapa:2025biy,Liu:2021zxr,Cho:2022syn,Levi:2020lfn,Levi:2022rrq,Mogull:2020sak,Jakobsen:2021smu,Jakobsen:2022psy,Driesse:2024xad,Cho:2022syn,Jakobsen:2022fcj,Jakobsen:2023ndj,Haddad:2024ebn,Driesse:2024feo,Hoogeveen:2025tew} and both methods mutually benefit from each other.

The QFT description of spinning black holes is more involved due to the intrinsic dynamics of angular momenta. However, progress has been made in several directions at both finite and infinite order in spin. A key insight is that the spin-multipole expansion of a Kerr black hole is described by heavy higher-spin particles combined with coherent states to convert the quantum spin into classical spin \cite{Damgaard:2019lfh,Aoude:2021oqj}. High-spin computations are made feasible by combining the effective field theory for heavy higher-spin particles, called Heavy Particle Effective Theory (HPET),  with on-shell methods~\cite{Aoude:2020onz}.
In HPET, the heavy particle momentum is decomposed into $p^\mu = m v^\mu +k^\mu$, and the mass $m$ is assumed to be much larger than the interaction scale $|k|$. A natural expansion in $|k|/m$ immediately follows, in which propagators linearize and numerators exponentiate.
With this new tool, we computed the one-loop conservative amplitude for two black holes to all spin orders~\cite{Aoude:2022thd,Aoude:2023vdk}, which resummed into Bessel-like functions. This result makes use of the minimal three-point amplitude~\cite{Arkani-Hamed:2017jhn,Guevara:2018wpp,Chung:2018kqs,Guevara:2019fsj,Aoude:2020onz,Aoude:2021oqj} and the spurious-pole-free Compton amplitude~\cite{Aoude:2022trd,Haddad:2023ylx}. The literature on amplitudes for Kerr black holes is growing rapidly~\cite{Maybee:2019jus,Chung:2018kqs,Guevara:2018wpp,Guevara:2019fsj,Arkani-Hamed:2019ymq,Chung:2020rrz,Guevara:2020xjx,Kosmopoulos:2021zoq,Bautista:2021wfy,Jakobsen:2021lvp,Liu:2021zxr,Chiodaroli:2021eug,Haddad:2021znf,Chen:2021kxt,Cristofoli:2021jas,Guevara:2021yud,Jakobsen:2022fcj,Alessio:2022kwv,Bern:2022kto,Menezes:2022tcs,Jakobsen:2022zsx,Cangemi:2022bew,Bautista:2022wjf,Bjerrum-Bohr:2023jau,Haddad:2023ylx,Aoude:2023fdm,Jakobsen:2023hig,Bjerrum-Bohr:2023iey,Cangemi:2023ysz,Gatica:2023iws,Luna:2023uwd,Cangemi:2023bpe,Azevedo:2024rrf,Alaverdian:2024spu,Bohnenblust:2024hkw,Correia:2024jgr,Haddad:2024ebn,Alessio:2025nzd,Bohnenblust:2025gir}.  
At finite order, two and three loops have been obtained up to $\mathcal{O}(S^4)$
~\cite{FebresCordero:2022jts,Akpinar:2024meg,Akpinar:2025bkt}. 

An important aspect of scattering amplitudes is resummation. In certain kinematics limits, one can resum infinite sets of diagrams and obtain improved results compared to naive perturbation theory. There are several examples of well-established resummations in QFT, including in the eikonal, Regge, soft, and collinear limits. All of them have been explored in scattering amplitudes for gravitational waves~\cite{DiVecchia:2021bdo,DiVecchia:2023frv,Rothstein:2024nlq}. While the Post-Newtonian (PN) expansion---a double expansion in $G$ and the velocity---is the relevant expansion for the inspiral phase of the two-body problem, the PM expansion resums the velocity dependence of the PN expansion, 
unveiling hidden structures in the expanded result. 

More recently, a resummation of certain loop topologies, known as fan diagrams, was understood to capture the geodesic motion of a probe particle in a background. Resumming these diagrams to all orders in $G$ reproduces the zeroth order in the self-force expansion~\cite{Cheung:2023lnj,Cheung:2024byb,Mougiakakos:2024nku,Mougiakakos:2024lif}. In the same spirit, when a closed form for all spin orders is present, hidden patterns are unveiled in simple, compact formulae, offering insight in the perturbative expansion.

In this work, we explore this resummation, called \textit{spin resummation}~\cite{Aoude:2022thd}, to high loop orders. We achieve this by considering the set of Feynman diagrams generated by the Kerr black hole background as \textit{Kerr generating functions} for multi-loop integrands. With the use of these functions, we express the amplitudes in a compact form to all orders in spin.
The spin-resummed results allow us to search for structure in the amplitudes, and we uncover a hidden simplicity for a tidal probe in a Kerr black hole background that extends to \textit{all loop orders}.

The paper is organized as follows. 
In sec.~\ref{sec:GeneratingFunctions}, we describe the relevant all-loop arbitrary tensor integrals, captured by the generating functions, relevant for scattering in a Kerr black hole background. Explicit forms of these functions are shown in the \cite{supp}.
In sec.~\ref{sec:NeutronStars}, we describe the relevant tidal operators used to model neutron stars.
The scattering of neutron stars in a Kerr background is shown in sec.~\ref{sec:Amplitudes}, exposing a hidden simplicity in the result. We conclude in sec.~\ref{sec:Conclusion}.



\section{Kerr Generating Functions}\label{sec:GeneratingFunctions}

In the probe limit, the Kerr black hole background is generated by three-point spin-exponentiated amplitudes. These amplitudes are not only fascinating in their own right, but they can be leveraged to perform the tensor reduction for multi-loop integrals. 
We achieve this by considering the set of integrands generated by the Kerr black hole background as \textit{Kerr generating functions} for multi-loop integrands. As such, we can efficiently perform the tensor reduction for a high-loop integrand---which would be unfeasible with traditional methods---simply by differentiating the generating functions with respect to the spin of the black hole. This work draws inspiration from~\cite{Feng:2022hyg} (see also~\cite{Chen:2024bpf,Hu:2024kch,Brandhuber:2024qdn}), and extends this method to all loop orders for scattering in a Kerr black hole background. Of course, this also captures the scattering in a Schwarzschild black hole background by sending the spin of the black hole to zero at the end of the computation.

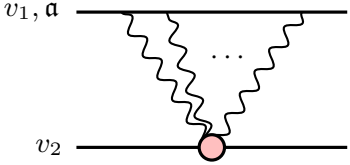
\begin{figure}[!htb]
	\centering
	\begin{tikzpicture}[scale=1.2, transform shape]
		\usetikzlibrary{decorations.pathmorphing}
		\tikzset{snake it/.style={decorate, decoration=snake}}
		\path [draw=black, very thick] (3.5,0.0) -- (6.5,0.0);
		\path [draw=black, very thick] (3.5,-1.5) -- (6.5,-1.5);
		\path [draw=black, snake it, thick] (5.0,-1.5) -- (4.0,0);
		\path [draw=black, snake it, thick] (5.0,-1.5) -- (4.5,0);
		\path [draw=black, snake it, thick] (5.0,-1.5) -- (6.0,0);
            \filldraw[fill=red!25,very thick] (5,-1.5) circle (4.0pt);
            \node (A2) at (5.2,-0.5) {$\cdots$};
		\node (A1) at (3.0,0.0) {$v_1, \fa$};
		\node (A2) at (3.2,-1.5) {$v_2$};	
	\end{tikzpicture}
        \caption{Multi-triangle diagram for the scattering of a neutron star in a black hole background. This class of diagrams serves as a generating function for  scattering in a Kerr black hole background.}   
        \label{fig:TwoLoopFeynmanDiag}
\end{figure}

The $L$-loop Kerr generating functions are
\begin{align}
	G^{(L,k)}(\mathfrak{a})  
	&\equiv \int \hat{\dd}^D\ell_1 \cdots    \hat{\dd}^D\ell_L
	\frac{1}{(2v_1\cdot \ell_1)\cdots (2v_1\cdot \ell_{1\cdots L})} \nonumber \\
	&\times \frac{1}{\ell_1^2\cdots \ell_L^2 (q-\ell_{1\cdots L})^2} \text{exp}[\ell_{1\cdots k}\cdot \mathfrak{a}] \,, 
\end{align}
where $\ell_{1\cdots n} = \ell_{1} + \ell_{2} + \dots + \ell_{n}$ and $\hat{\dd}^D\ell_i = \dd^D\ell_i/(2\pi)^D$. The index $k$ labels the helicity sector and $0\leq k\leq L+1$, $\mathfrak{a}$ is the ring radius of the Kerr black hole, $v_1$ is the velocity of the black hole, and $q$ is the transferred momentum. The relevant multi-triangle diagram is shown in \cref{fig:TwoLoopFeynmanDiag}.
This generating function describes any scattering in a Kerr black hole background.

We compute this integral recursively and obtain the compact expression
\begin{align}\label{eq:FgeneratingFunction}
    G^{(L,k)}(\mathfrak{a})  
    &= \sum_{i,j=0}^{\infty} c^{(L,k)}_{i,j} \frac{(Q/16)^{i}}{i!} \frac{(q\cdot \mathfrak{a}/2)^{j}}{j!} I_{L\triangle} \,,
\end{align}
where $Q = (q\cdot\mathfrak{a})^2 - q^2 \mathfrak{a}^2$ and $I_{L\triangle}$ is the $L$-loop scalar multi-triangle integral
\begin{align}
    \label{eq:LScalarFan}
	I_{L\triangle} 
	&= \frac{1}{2^L(L+1)!}
	\frac{\Gamma[\bm{d}]^{L+1}\Gamma[1-\bm{d}L]}{(4\pi)^{(\bm{d}+1)L}\Gamma[(L+1)\bm{d}]} (\bm{q}^2)^{\bm{d}L-1}
\end{align}
where $\bm{d} = (D- 3)/2$.
We have dropped other master integrals that do not have the appropriate pole structure to contribute to the classical limit. \Cref{eq:FgeneratingFunction} is useful because we have a closed-form expression for the coefficients which depend on the loop order $L$ and the $k^{\textrm{th}}$ helicity sector of the scattering amplitude. The coefficients are
\begin{align}
    c_{i,j}^{(L,k)} =& \frac{2^{2i+j} \left(\bm{d}(L+1-k)\right)_{i} \left(\bm{d}k\right)_{i+j}}{\left(\bm{d}L \right)_{i} \left(\bm{d}(L+1)\right)_{2i+j}} \\
    &\times \,_3F_2\left[\begin{Bmatrix}\bm{d}(L+1-k) + i, \tfrac{1-j}{2}, -\tfrac{j}{2} \\  \bm{d}L + i, 1 - \bm{d}k - i - j \end{Bmatrix}, 1\right] \,, \notag
\end{align}
where $(x)_n$ is the rising Pochhammer symbol.
The coefficients for the $k=1$ sector simplify to
\begin{align}
    c_{i,j}^{(L,1)} &= \frac{1}{\left(\bm{d}+1/2\right)_{i}} \frac{(2\bm{d})_{2i+j}}{\left(\bm{d}(L+1)\right)_{2i+j}} \,.
\end{align}
One can directly notice the simplicity at one-loop order, where the second fraction cancels out. 

All tensor integrals can be generated by taking derivatives with respect to $\mathfrak{a}$. The spin of the black hole satisfies the spin supplementary condition, $v_1 \cdot \mathfrak{a} = 0$. Because of this, the spin derivatives project the tensor result onto the three-dimensional space orthogonal to $v_1$. Some important properties of the Kerr generating functions are that they vanish under the following operations
\begin{align}
    \partial_{\mathfrak{a}} \cdot \partial_{\mathfrak{a}} \left( G^{(L,1)}(\mathfrak{a}) \right) =
    0 = v_1 \cdot \partial_{\mathfrak{a}} \left( G^{(L,k)}(\mathfrak{a}) \right) 
\end{align}
up to other master integrals which are irrelevant for our classical computations.

As a first application of these generating functions, we consider a tidal probe in a Kerr black hole background.
To organize our expressions, we define the operator
\begin{align}\label{eq:DiffOperator}
    &\mathds{D}_{(a,b,c,d)} 
    \equiv \left[ 2\frac{ \left(v_2\cdot \partial_\mathfrak{a} \right)^{2}}{4 q^2} + (2\gamma^2-1)\frac{\nabla_{\pm} \cdot \partial_{\mathfrak{a}}}{4q^2} \right]^{a} \\
    &\left[ \gamma^2 \frac{ \left(v_2\cdot \partial_\mathfrak{a} \right)^{2}}{4 q^2}  {+} \gamma^2 (\gamma^2-1)\left( \frac{\partial^2_{\mathfrak{a}}}{4q^2} {-} \frac{(q \cdot \partial_{\mathfrak{a}})^2}{4q^4}\right)\right]^{b}  \left[ \frac{\nabla^2_{\pm}}{4q^2} \right]^{c} \left[ \frac{\partial^2_{\mathfrak{a}}}{4q^2} \right]^{d} \notag 	
\end{align}
where $\nabla_{\pm}^\mu = \partial_\fa^\mu \mp 2 q^{\mu}$ and $\gamma = v_1 \cdot v_2$.  The differential operators act on the Kerr generating functions to form
\begin{align}
	K^{(L,k)}_{(a,b,c,d)} &= \sum_{\pm} e^{\mp q \cdot \mathfrak{a}} 
	\mathds{D}_{(a,b,c,d)} G^{(L,k)}(\pm 2\mathfrak{a}) \,,\\
    \tilde{K}^{(L,k)}_{(a,b,c,d)} &= i\gamma\varepsilon(v_1 v_2 q \partial_{\mathfrak{a}} )
    K^{(L,k)}_{(a,b,c,d)} \,,
\end{align}
which are relevant for the even-in-spin and odd-in-spin amplitudes, respectively. The Levi-Civita symbol appears because the spin vector is defined in $D=4$. 
For the $k=0$ sector, the amplitudes are expressed in terms of
\begin{align}\label{eq:Kdefk0}
    K^{(L,0)} &\equiv \frac{1}{4}\cosh(q\cdot \fa) \\
    &\times \lim_{\fa \rightarrow 0} \left( K^{(L,1)}_{(0,0,L+1,0)} - 2 K^{(L,1)}_{(0,0,L,0)} + K^{(L,1)}_{(0,0,L-1,0)} \right) \,. \notag
\end{align}
For the tidal probe, we need only a limited set of these functions. For example, for the one- and two-loops result, we need $K^{(1,1)}_{(a,b,0,0)}$ and $K^{(2,1)}_{(a,b,1,0)}$ up to $(a+2b)=2$, respectively. In \cite{supp}, we list the relevant $K_{(a,b,L-1,0)}^{(L,1)}$ functions used for the NMHV amplitudes in this work. In the simple case of $L=1$ and $k=1$, we recover the hypergeometric function obtained in~\cite{Aoude:2022thd,Aoude:2023vdk} for the scattering of two black holes at one-loop order but to all orders in spin;
\begin{align}\label{eq:oneLoopHypergeometric}
    K^{(1,1)}_{(0,0,0,0)} 
    &=\,_0\tilde{F}_{1}\left(1; \frac{Q}{4} \right)\,,
\end{align}
where $\,_0\tilde{F}_{1}$ is the regularized confluent hypergeometric function, which is closely related to Bessel functions. Further spin-derivatives will lead to the function $\,_0\tilde{F}_{1}\left(n+1; \frac{Q}{4} \right)$.
In this case, the dependence on $(q\cdot \mathfrak{a})$ is exactly cancelled from the generating functions, leaving only the dependence on $Q$. Thus, the \textit{shift symmetry} of black hole scattering at one loop is manifest \cite{Aoude:2022trd,Bern:2022kto}. At $L$ loops, the Kerr generating functions can be written as Kamp\'e de F\'eriet generalized hypergeometric functions.



\section{Neutron stars}
\label{sec:NeutronStars}

We model the neutron star by considering tidal operators constructed out of the electric and magnetic components of the gravitational field,
\begin{align}
	E_{\mu\nu} = C_{\mu\alpha\nu\beta}v^{\alpha}v^{\beta}, \quad 
	B_{\mu\nu} = \frac12 \epsilon_{\alpha\beta\gamma\mu} C^{\alpha\beta}_{\quad\delta\nu}v^{\gamma}v^{\delta}  \,,
\end{align}
where $C_{\mu\nu\rho\sigma}$ is the Weyl tensor and $v^\mu$ is the four-velocity of the neutron star. The complete infinite tower of tidal operators for scalar particles was obtained in~\cite{Haddad:2020que} and for spin-$\frac12$ in~\cite{Aoude:2020ygw}. Here, for the $L$-loop computation, we will focus on the tidal operator built from the electric component of the Weyl tensor,
\begin{align}
\label{eq:TidalOperators}
    \mathcal{O}^{(L)}_{\rm tidal} 
    &=c_{E^{L+1}} \varphi^2\, \textrm{tr} [E^{L+1}] \\\
    &\equiv c_{E^{L+1}} \varphi^2 E_{\mu_1 \mu_2}E^{\mu_2\mu_3}E_{\mu_3\mu_4}\cdots E^{\mu_{L+1}\mu_1}\notag.
\end{align}
The results for general tidal operators with both electric and magnetic components of the Weyl tensor are presented in \cite{supp}.
These are the leading non-linear tidal operators for neutron stars. These tidal operators are not special per say, and they will mix with other tidal operators at higher loops~\footnote{We thank the referee for discussions on this point.}. Nevertheless, they will serve as a testing ground for the application of the Kerr generating functions.



\section{Scattering of Kerr black holes and neutron stars}\label{sec:Amplitudes}

Using the Kerr generating functions, we compute the leading scattering of a neutron star with mass $m_2$, velocity $v_2$, and non-linear tidal operator $\varphi^2 R^{L+1}$ in a Kerr black hole background with mass $m_1$, velocity $v_1$, and ring radius $\mathfrak{a}$. 
In this work, we present the one-, two-, three-, and four-loop results for the operators given in~\cref{eq:TidalOperators}. The result for general tidal operators is given in \cite{supp}. 

As we will see, the leading effects of these tidal operators share a remarkably similar structure, which is obscured without the use of the Kerr generating functions.
These results extend the computations in \cite{Bern:2020uwk}, which considered the same leading non-linear tidal operators, but in a Schwarzschild background.
It is convenient to split the $L$-loop amplitude into even and odd powers in spin:
\begin{align}\label{eq:AmpLloopSplit}
    \cM_{\varphi^2 R^{L+1}} {=} 2 m_1c_{E^{L+1}}\left(\frac{\kappa^2 m_1\, q^2}{16}\right)^{\!\!L+1}\!\!  \left( \cM_{\varphi^2 R^{L+1}}^{\rm even} {+} \cM_{\varphi^2 R^{L+1}}^{\rm odd} \right) \,.
\end{align}
The results for the other the general tidal operators are obtained from simple shifts of Wilson coefficients of the results for $\varphi^2 E^{L+1}$, as described in \cite{supp}. Of course, the full $L$-loop amplitude contains pieces beyond the ones we consider here, such as the insertion of multiple tidal operators and high-order corrections to the leading-order results for each tidal operator. 

We compute the amplitude from on-shell unitarity cuts and sum over the helicities of the exchanged gravitons. Because of this, we can further decompose the even-in-spin and odd-in-spin amplitudes into the different helicity sectors. As we will see, this decomposition is useful since the different helicity sectors of the amplitude expose hidden structure in the final result. In the case when all internal gravitons have the same helicity, say positive helicity, we call it the Maximally Helicity Violating (MHV) amplitude. Correspondingly, the amplitude with mostly positive and $k$ negative helicity gravitons, we name the N$^{k}$MHV amplitude. At $L$ loops, the sum over the helicity sectors is
\begin{align}
	\cM_{\varphi^2 R^{L+1}}^{\rm even} &= \sum_{k=0}^{\floor{L/2}} \cM_{\varphi^2 R^{L+1}}^{{\rm even},(k)} \,,
\end{align}
and similar for the odd-in-spin amplitude. 
For example, at one-loop order we have $\cM_{\varphi^2 R^2}^{{\rm even},(0)}  = \cM_{\varphi^2 R^2}^{\rm even,++} + \cM_{\varphi^2 R^2}^{\rm even,--}$ and $\cM_{\varphi^2 R^2}^{{\rm even},(1)}  = \cM_{\varphi^2 R^2}^{\rm even,+-} + \cM_{\varphi^2 R^2}^{\rm even,-+}$. The counting of the helicity sectors follows the same pattern as the self-force expansion in that we only need one new ingredient for every second loop order. At tree-level, only the MHV amplitude contributes. At one and two loops, the NMHV amplitude is also needed. For three and four loops, we also need the N$^{2}$MHV amplitude, etc.

The MHV amplitudes are particularly simple. The odd-in-spin amplitude vanishes, while the even-in-spin amplitude is fully captured by a hyperbolic cosine stemming from the spin-exponentiated three-point amplitudes. 
The $L$-loop MHV amplitudes are
\begin{align}\label{eq:AmpEvenMHV}
	\cM_{\varphi^2 R^{L+1}}^{\rm even,(0)} 
	=& c_{L} K^{(L,0)} \,,
\end{align}
where $c_L$ is given in Table~\ref{tab:OverallCoefficients} and $K^{(L,0)}$ was defined in~\cref{eq:Kdefk0}. For example, at one loop the result is $K^{(1,0)}=1/2\cosh(q\cdot \fa) I_{1\triangle}$. Given the integrand structure, the MHV amplitude is independent of $\gamma$, making it subleading in the high-energy limit, where $\gamma \to \infty$. 

Moving to the NMHV amplitudes, the loop integrand will now contain an exponential factor of the loop integration variable. Thus, we need to tensor reduce an infinite number of tensor integrals. Of course, this is precisely captured by the Kerr generation functions. 
The $L$-loop NMHV amplitudes are simply
\begin{align}
	 \cM_{\varphi^2 R^{L+1}}^{\rm odd,(1)} 
	=& 4 d_{L} \tilde K^{(L,1)}_{(1,0,L-1,0)} \,, \\
    \cM_{\varphi^2 R^{L+1}}^{\rm even,(1)} 
    =& d_{L} \bigg\{ K^{(L,1)}_{(2,0,L-1,0)}  - 4 K^{(L,1)}_{(0,1,L-1,0)} \bigg\} \,.
\end{align}
Lastly, the $L$-loop N$^2$MHV amplitudes are
\begin{align}
	 \cM_{\varphi^2 R^{L+1}}^{\rm odd,(2)} 
	=& -e_{L} \left\{\frac{1}{2}\tilde K^{(L,2)}_{(3,0,L-3,0)} - 2 \tilde K^{(L,2)}_{(1,1,L-3,0)}  \right\} \notag \\
	&- 6 \left(f_{L} - \frac{13}{108} e_{L} \right) \tilde K^{(L,2)}_{(1,0,L-2,1)}  \,,
\end{align}
and
\begin{align}
    &\cM_{\varphi^2 R^{L+1}}^{\rm even,(2)} 
    = + g_{L} K^{(L,2)}_{(0,0,L-1,2)}  \\
    &\quad + e_{L} \bigg\{ \frac{1}{8} K^{(L,2)}_{(4,0,L-3,0)} - 3 K^{(L,2)}_{(2,1,L-3,0)} 
    + 2 K^{(L,2)}_{(0,2,L-3,0)} \bigg\} \notag \\
    &\quad + 3\left(f_{L} - \frac{13}{108} e_{L}\right) \bigg\{  K^{(L,2)}_{(2,0,L-2,1)}  - 4 K^{(L,2)}_{(0,1,L-2,1)}  \bigg\} \,. \notag
\end{align}
%
%
The overall coefficients $c_L$, $d_L$, $e_L$, $f_L$, and $g_{L}$ are
\begin{center}
\label{tab:OverallCoefficients}
\renewcommand{\arraystretch}{1.5}
\begin{tabular}{ |c|c|c|c|c|c| } 
	\hline
	Loop $L$ & $c_{L}$ & $d_{L}/c_{L}$ & $e_{L}$ & $f_{L}/e_{L}$ & $g_{L}/e_{L}$ \\
	\hline
	1 & 8 & 1 & -  & - & - \\ 
	2 & 3 & 3 & - & - & - \\ 
	3 & $\frac{16}{35}$ & 4 & $\frac{27}{16}$ & $\frac{16}{27}$ & $\frac{227}{24}$ \\ 
	4 & $\frac{1}{42}$ & 5 & $\frac{120}{77}$ & $\frac{61}{108}$ & $\frac{7}{8}$ \\
	\hline
\end{tabular}
\end{center}
In the spinless limit, we find perfect agreement with the literature~\cite{Bern:2020uwk}.

With these results in hand, we can make several observations. First, the amplitude is written in a remarkably simple form, with the spin-dependence of the 
black hole resummed to all orders. These are the first higher-loop results of \textit{spin resummation} for processes involving neutron stars. Previous one-loop results for spin resummation of black hole scattering are given in~\cite{Aoude:2022thd,Aoude:2023vdk}. 

Second, the shift symmetry conjectured for black holes in~\cite{Aoude:2022trd,Bern:2022kto} is broken by the presence of a tidal probe in general. However, at one loop the sole culprit is the MHV even-in-spin amplitude. By using the results in \cite{supp} and fixing the Wilson coefficients to the values $c_{B^2} = c_{E^2}$, we recover the shift symmetry for the full one-loop amplitude. The result is then expressed in terms of the same hypergeometric function, \cref{eq:oneLoopHypergeometric}, which enters in the analogous result for black holes. 

Lastly, by fixing the Wilson coefficients to special values, the
leading terms in the high-energy limit cancel and thus the amplitude exhibits an improved high-energy behavior. Ref.~\cite{Bini:2020flp} showed that with the choice of
coefficients $c_{B^2} = - c_{E^2}$ at one loop one obtains the Kretschmann scalar $\mathcal{O}_{\rm Kr}^{(1)} \equiv {\rm tr}[E^2]-{\rm tr}[B^2]$, which is invariant under boosts. In general, it is possible to tune the coefficients such that the odd-in-spin amplitude vanishes and the even-in-spin amplitude becomes independent of $\gamma$. These operators generalize the Kretschmann scalar up to four loops:
\begin{align*}
	\mathcal{O}^{(1)}_{\rm Kr} =& {\rm tr}[E^2]-{\rm tr}[B^2] \,, \\
	\mathcal{O}^{(2)}_{\rm Kr} =& {\rm tr}[E^3]-3{\rm tr}[EB^2] \,, \\
	\mathcal{O}^{(3)}_{\rm Kr} =& {\rm tr}[E^4] + {\rm tr}[B^4] - 4 {\rm tr}[E^2B^2] - 2 {\rm tr}[EBEB] \,, \\
	\mathcal{O}^{(4)}_{\rm Kr} =& {\rm tr}[E^5] + 5 {\rm tr}[EB^4] - 5 {\rm tr}[E^3B^2] - 5 {\rm tr}[EBEBE] \,. 
\end{align*}
These operators are invariant under boosts because they do not have any dependence on the velocity of the neutron star. Moreover, they are nonzero only in the MHV sector. Our amplitude-based computations make both these properties manifest.



\section{Eikonal Phase}
\label{sec:ScatteringAngle}

To make use of our all-spin amplitudes, we compute the eikonal phase defined as
\begin{align}\label{eq:eikonal}
	\chi_{\varphi^2 R^{L+1}}^{(k)} {=} \frac{1}{4m_1m_2\sqrt{\gamma^2-1}}\int \! \frac{\dd^{D-2} \bm{q}}{(2\pi)^{D-2}} e^{i\bm{b}\cdot \bm{q}} \cM_{\varphi^2 R^{L+1}}^{(k)}.
\end{align}
By focusing on the tidal scattering of the generalized Kretschmann scalars defined above, our computation simplifies dramatically; the spin dependence is fully captured by $\cosh(q\cdot \mathfrak{a})$. We exploit this fact by writing the Kretschmann eikonal phase in an alternate form linked to the Newman-Janis shift~\cite{Newman:1965tw};
\begin{align}{}
	\chi_{\textrm{Kr}} {=} \frac{1}{2}\sum_{\pm} \chi_{\varphi^2 R^{L+1}}^{(k=0,s=0)}(\bm{b}\mp i \mathfrak{a}) \,,
\end{align}
where $\chi_{\varphi^2 R^{L+1}}^{(k=0,s=0)}(\bm{b})$ is the MHV eikonal phase defined in \cref{eq:eikonal} in the spinless limit.
The spinless MHV eikonal phase in $D=4$ is
\begin{align}
	&\chi_{\varphi^2 R^{L+1}}^{(k=0,s=0)} 
	{=} 
	\frac{c_{E^{L+1}}c_L(\kappa^2m_1^2/16)^{L+1}}{4m_1 m_2\sqrt{\gamma^2-1}} 
	\left(\frac{1}{{\pi \bm b}^2}\right)^{\tfrac{3L}{2}+1}  \\
    &\qquad\,\,\,\,\times
    \frac{(-1)^{L+1}\pi^{L/2}\Gamma[\tfrac{3L}{2}+1]^2}{3\Gamma[L]\Gamma[L+2]}\left(\delta_{L,1} + \frac{4}{(9L^2-1)} \right) \,. \notag
\end{align}
The MHV eikonal is obtained from the amplitude in \cref{eq:AmpEvenMHV} together with the overall factors in \cref{eq:AmpLloopSplit}.


\section{Conclusion}
\label{sec:Conclusion}

We derived the Kerr generating functions, which describe any scattering in a Kerr black hole background with the spin resummed into compact expressions. As a first application of these generating functions, we computed the leading nonlinear tidal deformations of a neutron star in a Kerr black hole background up to 4 loops. As a by-product of this computation, we uncovered a hidden simplicity in the final result, where the spin-resummed amplitudes at various loop orders in a given helicity sector are captured by the \textit{same} Kerr generating functions, a property that extends to all loop orders.

High-loop, high-spin computations are difficult due to the complexity in the computations. Using the traditional pipeline of integration-by-parts identities to reduce the integrands encountered into master integrals quickly becomes unfeasible for high-spin computations. Here we present an alternative. By expressing the integrands in terms of the Kerr generating functions, we have simplified the problem into the problem of obtaining these generating functions. At leading order in the self-force expansion, these generating functions are fully captured by \cref{eq:FgeneratingFunction} to all loop orders. We expect that this new organization of the computation will also make the full computation of the scattering of two black holes at higher loops and higher spin orders trackable. We leave this for future work.

\vspace*{.2cm}

\begin{acknowledgments}

We thank Kays Haddad and Franz Herzog for useful discussions.
We also thank Jung-Wook Kim for valuable comments on a previous version of this draft.
R.A. is supported by UK Research and Innovation (UKRI) under the UK government’s
Horizon Europe Marie Sklodowska-Curie funding guarantee grant [EP/Z000947/1]
and by the STFC grant “Particle Theory at the Higgs Centre”. Some manipulations were performed with the help of \texttt{FeynCalc}~\cite{Kublbeck:1992mt,Shtabovenko:2016sxi,Shtabovenko:2020gxv}.
\end{acknowledgments}

\twocolumngrid
\bibliographystyle{apsrev4-2}
\bibliography{AllLoopTidal_v2}

\clearpage
\newpage
\appendix

\onecolumngrid

\begin{center}
	\textbf{\large Supplemental Material for ``Hidden simplicity in the scattering for neutron stars and black holes"}\\[.2cm]
	\vspace{0.05in}
	{Rafael Aoude and Andreas Helset}
\end{center}

\section{Properties of the Kerr generating functions}
The generating function in Eq.\eqref{eq:FgeneratingFunction} has interesting properties when operated with \eqref{eq:DiffOperator}. Mainly, it vanished when
\begin{align}
v_2 \cdot \partial_{\mathfrak{a}} \left( G^{(L,k)}(\mathfrak{a}) \right) = 0.
\end{align}
since it pinches one of the massive propagators. For specific values of $k$, we can also write more properties
\begin{align}
    k=1 &\qquad\ \partial_\fa^2 
    \left( G^{(L,1)}(\mathfrak{a}) \right) = 0\\
    k=L &\qquad (q-\partial_\fa)^2
    \left( G^{(L,L)}(\mathfrak{a}) \right) = 0
\end{align}

\section{Explicit Tensor Integrals}
Here we list some of the relevant $K^{(L,k)}_{(a,b,c,d)}$ functions used in this work.  For the NMHV  $(k=1)$ results, we will need the following even-in-spin structures
\begin{align}
	K^{(L,1)}_{(2,0,L-1,0)}  &=
	\sum_{i,j=0}^{\infty} 
	\bigg[\frac14 
	\left(
	\frac{3(\gamma^2-1)^2}4 \braket{c^{(L,1)}_{i+2,j}}_{L-1}  -
	(2\gamma^2-1)(\gamma^2-1)\braket{c^{(L,1)}_{i+1,j+1}}_{L-1}    +(2\gamma^2-1)^2\braket{c_{i,j+2}^{(L,1)}}_{L-1} 
	\right)\\
	&\qquad\quad+\frac{V}{8}\left(
	\frac{(\gamma^2-1)}{2}\braket{c^{(L,1)}_{i+3,j}}_{L-1} 
	-(2\gamma^2-1)\braket{c^{(L,1)}_{i+2,j+1}}_{L-1} 
	\right) +\frac{V^2}{64} \braket{c_{i+4,j}^{(L,1)}}_{L-1} \bigg]
	\frac{(Q/4)^i f_j(q\cdot \fa) }{i!j!} I_{L\triangle}
	\notag\\
	K^{(L,1)}_{(0,1,L-1,0)} &=
	\sum_{i,j=0}^{\infty} 
	\bigg[
	\frac{\gamma^2(\gamma^2-1)}{8}
	\left(\braket{c_{i+1,j}^{(L,1)}}_{L-1}-    
        2\braket{c_{i,j+2}^{(L,1)}}_{L-1}
	\right)+ \gamma^2\frac{V}{16} c_{i+2,j}^{(L,1)}
	\bigg]
	\frac{(Q/4)^i f_j(q\cdot \fa) }{i!j!} I_{L\triangle}
\end{align}
and for the odd-in-spin
\begin{align}
	\tilde K^{(L,1)}_{(1,0,L-1,0)} =
	\frac{i\gamma q^2\cE_1}{2}
	\sum_{i,j=0}^{\infty} 
	\left[\frac{(\gamma^2-1)}{4}\braket{c_{i+2,j}^{(L,1)}}_{L-1}
	   -\frac{(2\gamma^2-1)}{2}\braket{c_{i+1,j+1}^{(L,1)}}_{L-1}
	+\frac{V}{8} \braket{c_{i+3,j}^{(L,1)}}_{L-1}
	\right]
	\frac{(Q/4)^i f_j(q\cdot \fa) }{i!j!} I_{L\triangle}
\end{align}
where $\cE_{1} \equiv \epsilon^{\mu\nu\alpha\beta}v_{1\mu}v_{2\nu}q_{\alpha}\mathfrak{a}_{\beta}$ and $V\equiv q^2(v_2\cdot \fa)$, and defined some compact notation
\begin{align}
\braket{c_{i,j}^{(L,k)}}_n = \sum_{r=0}^n \binom{n}{r} (-1)^r
c_{i,j+r}^{(L,k)},
\qquad
f_j(q\cdot \fa)  =  \sum_\pm e^{\mp q\cdot \fa} (\pm q\cdot \fa)^j.
\end{align} 

\section{Results for General Tidal Operators}

In the main text, we present the one-, two-, three-, and four-loop results for the tidal operator built from the electric component of the Weyl tensor. Here, we present the complete result for all independent tidal operators built from both the electric and magnetic components of the Weyl tensor. Note that the number of independent tidal operators matches the number of coefficients in the full amplitude. At one and two loops, we have two tidal operators and the coefficients $c_L$ and $d_L$. At three and four loops, we have five tidal operators and the coefficients $c_L$, $d_L$, $e_L$, $f_L$, and $g_L$. This correspondence between the operators and the amplitude extends to all loop orders.

\subsection*{1 Loop}

The tidal operators are
\begin{align}
\label{eq:TidalOperators1Loop}
    \mathcal{O}^{(1)}_{\rm tidal} 
    &=c_{E^{2}} \varphi^2\, \textrm{tr} [E^{2}] + c_{B^{2}} \varphi^2\, \textrm{tr} [B^{2}] \, .
\end{align}
The full scattering result is obtained via the following shifts:
\begin{align}
	c_{E^2} c_{L} &\rightarrow (c_{E^2} - c_{B^2}) c_{L} \,, \\
	c_{E^2} d_{L} &\rightarrow (c_{E^2} + c_{B^2}) d_{L} \,.
\end{align}
\subsection*{2 Loop}
The tidal operators are
\begin{align}
\label{eq:TidalOperators2Loop}
    \mathcal{O}^{(2)}_{\rm tidal} 
    &=c_{E^{3}} \varphi^2\, \textrm{tr} [E^{3}] + c_{EB^{2}} \varphi^2\, \textrm{tr} [EB^{2}] \, .
\end{align}
The full scattering result is obtained via the following shifts:
\begin{align}
	c_{E^3} c_{L} &\rightarrow (c_{E^3} - c_{E B^2}) c_{L} \\
	c_{E^3} d_{L} &\rightarrow \left( c_{E^3} + \frac{1}{3} c_{E B^2} \right) d_{L} 
\end{align}

\subsection*{3 Loop}

The tidal operators are
\begin{align}
\label{eq:TidalOperators3Loop}
    \mathcal{O}^{(3)}_{\rm tidal} 
    &=c_{E^{4}} \varphi^2\, \textrm{tr} [E^{4}] + c_{B^{4}} \varphi^2\, \textrm{tr} [B^{4}] + c_{E^2B^2} \varphi^2\, \textrm{tr} [E^2B^2] + c_{EBEB} \varphi^2\, \textrm{tr} [EBEB] + c_{(E^2)(B^2)} \varphi^2\, \textrm{tr} [E^2]\, \textrm{tr} [B^2] \, .
\end{align}
The full scattering result is obtained via the following shifts: 
\begin{align}
	c_{E^4} c_{L} &\rightarrow \left(c_{E^4} + c_{B^4} - c_{E^2B^2} - c_{EBEB} - 2 c_{(E^2)(B^2)}\right) c_{L} \,, \\
	c_{E^4} d_{L} &\rightarrow \left(c_{E^4} - c_{B^4}\right) d_{L} \,, \\
	c_{E^4} e_{L} &\rightarrow \left(c_{E^4} + c_{B^4} + c_{E^2B^2} - c_{EBEB} + 2 c_{(E^2)(B^2)}\right) e_{L} \,, \\
	c_{E^4} f_{L} &\rightarrow \left(c_{E^4} + c_{B^4} + c_{EBEB} + 2 c_{(E^2)(B^2)}\right) f_{L} \,, \\
	c_{E^4} g_{L} &\rightarrow \left(c_{E^4} + c_{B^4} + \left(\frac{41}{681}\right) c_{E^2B^2} + \left(1 - \frac{82}{681}\right) c_{EBEB} + \left(-1 - \frac{5}{681}\right) c_{(E^2)(B^2)}\right) g_{L} \,.
\end{align}

\subsection*{4 loop}

The tidal operators are
\begin{align}
\label{eq:TidalOperators4Loop}
    \mathcal{O}^{(4)}_{\rm tidal} 
    &=c_{E^{5}} \varphi^2\, \textrm{tr} [E^{5}] + c_{EB^{4}} \varphi^2\, \textrm{tr} [EB^{4}] + c_{E^3B^2} \varphi^2\, \textrm{tr} [E^3B^2] + c_{EBEBE} \varphi^2\, \textrm{tr} [EBEBE] + c_{(E^2)(EB^2)} \varphi^2\, \textrm{tr} [E^2]\, \textrm{tr} [EB^2] \, .
\end{align}
The full scattering result is obtained via the following shifts:
\begin{align}
	c_{E^5} c_{L} &\rightarrow \left(c_{E^5} + c_{EB^4} - c_{E^3B^2} - c_{EBEBE} - \frac{6}{5} c_{(E^2)(EB^2)} \right) c_{L} \,, \\
	c_{E^5} d_{L} &\rightarrow \left(c_{E^4} - \frac{3}{5} c_{EB^4} - \frac{1}{5} c_{E^3B^2} - \frac{1}{5} c_{EBEBE} - \frac{6}{25} c_{(E^2)(EB^2)} \right) d_{L} \,, \\
	c_{E^5} e_{L} &\rightarrow \left(c_{E^4} + \frac{1}{5} c_{EB^4} + \frac{3}{5} c_{E^3B^2} - \frac{1}{5} c_{EBEBE} +  \frac{2}{5} c_{(E^2)(EB^2)}  \right) e_{L} \,, \\
	c_{E^5} f_{L} &\rightarrow \left(c_{E^4} + \frac{1}{5} c_{EB^4} + \frac{1}{5}\left( 1 - \frac{10}{61} \right) c_{E^3B^2} + \frac{1}{5}\left( 1 + \frac{10}{61} \right) c_{EBEBE} +  \frac{2}{5} c_{(E^2)(EB^2)}  \right) f_{L} \,, \\
	c_{E^5} g_{L} &\rightarrow \left(c_{E^4} + \frac{1}{5} c_{EB^4} - \frac{1}{5}\left(1 + \frac{4}{21}\right) c_{E^3B^2} + \frac{3}{5}\left(1 + \frac{4}{63}\right) c_{EBEBE} + \frac{1}{5}\left(-4 + \frac{26}{63}\right) c_{(E^2)(EB^2)} \right) g_{L} \,.
\end{align}

\end{document}